\documentclass[twocolumn,footinbib,showpacs]{revtex4-1}
\usepackage{graphicx}
\usepackage{amsmath}

\begin{document}

\title{Opinion formation model for markets with a social temperature and fear}
\author{Sebastian M.\ Krause}
\email{sebastian.krause@itp.uni-bremen.de}
\author{Stefan Bornholdt}
\email{bornholdt@itp.uni-bremen.de}
\affiliation{Institut f\"ur Theoretische Physik, Universit\"at Bremen, D-28359 Bremen, Germany}
\begin{abstract}
In the spirit of behavioral finance, we study the process of opinion formation among investors using a variant of the 2D Voter Model with a tunable social temperature. Further, a feedback acting on the temperature is introduced, such that social temperature reacts to market imbalances and thus becomes time dependent. In this toy market model, social temperature represents nervousness of agents towards market imbalances representing speculative risk. We use the knowledge about the discontinuous Generalized Voter Model phase transition to determine critical fixed points. The system exhibits metastable phases around these fixed points characterized by structured lattice states, with intermittent excursions away from the fixed points. The statistical mechanics of the model is characterized and its relation to dynamics of opinion formation among investors in real markets is discussed. 
\end{abstract}
\pacs{89.65.-s, 05.50.+q, 05.65.+b, 64.60.De}
\maketitle

\section{Introduction}
In the growing field of behavioral finance, market participants are characterized using psychological and sociological insight in order to go beyond the limited, yet common assumption of idealized rational investors \cite{sornette2004}. One major social effect among investors is the process of opinion formation. Opinion formation models as, for example, the voter model are broadly discussed in the physics literature, with special emphasis on emergent behavior including phase transitions \cite{castellano2009}. In this article, we discuss opinion formation in the context of markets by relating the voter model to a class of minimalistic market models inspired by models of statistical mechanics. 

A particularly simple toy model for markets is inspired by the Ising model with spins on a regular 2D lattice with a feedback, where the global magnetization controls the switching probabilities of single agents \cite{bornholdt2001}. It exhibits stylized facts typical for financial markets \cite{kaizoji2002,yamamoto2010} as broadly distributed returns \cite{mandelbrot1963} and volatility clustering \cite{gopikrishnan1999}, and can be related to macroscopic market models \cite{krause2011}. We here take this model as a starting point and show that the dynamical mechanism ruling the spin model can be mapped in an intuitive way onto a 2D voter model variant having a tunable social temperature. With the opinions ``buy'' and ``sell'', opinion polarization appears as market imbalance. We introduce a feedback representing the nervousness of agents: increasing market imbalance leads to increasing  temperature. 

The system shows time dependent market imbalance and temperature. We identify two alternating metastable phases. Phases of low temperature dominate the time series. In contrast to ordered phases in other systems, striped lattice states allow for a system of low polarization (each of the two opinions occupies a stripe which spans the lattice in one direction and is about the half of the lattice length in the other direction). These states  evolve due to undirected diffusion, finally leading to market imbalances resulting in a critical temperature. Metastable phases with fluctuations around the critical temperature enable the system to self-organize to balanced striped states. Using a mean-field Langevin description, we find two attracting critical fixed points at specific values of market imbalance and understand fluctuations around them. By considering different functions for the feedback, we find large fluctuations with consecutive melting and coarsening periods, or alternatively small fluctuations with critical lattice states, both leading to long range correlations and striped states. 

In order to connect the market model to the social systems perspective, we pick up a central line of research in the field of opinion formation. The paradigmatic voter model (VM) on regular lattices \cite{clifford1973} describes agents with two possible opinions, denoted as spin values $\pm 1$. A randomly chosen agent, with a certain probability $p$, adopts the opinion of one randomly chosen nearest neighbor. The universality class of the generalized voter model (GVM) \cite{dorn01} is characteristic for systems with parametrized interactions which show a special non-equilibrium phase transition in the presence of absorbing states (\cite{odor04,hin10} provide an overview of non-equilibrium phase transitions). These systems exhibit an abrupt phase transition (with a jump in the order parameter), however, show critical divergences \cite{dorn01}. The knowledge about this phase transition was extended using backward Fokker-Planck equations, mean field calculations, and a Langevin description \cite{cast09,ham05,vazquez2008}, and the time dependent coarsening process is well known \cite{dorn01}. The drift forces constituting the phase transition were described to act likewise even for finite systems \cite{krause12}. Similar opinion formation models were investigated for striped metastable states \cite{chen2005}, interfaces \cite{bordogna2011} as well as the influence of time dependent switching rates \cite{stark2008}. 

In the following we will formulate an opinion formation toy model for financial markets using an interesting representative of the GVM phase transition including the VM at its critical point \cite{oliv93,dorn01}, recently called the Group Voter Model (GRVM) \cite{krause12}. 
In Sec.\ \ref{sec:constant} we find different market modes for constant market temperature. 
In Sec.\ \ref{sec:feedback}, results for our artificial market with feedback are presented and the two metastable phases are discussed. 
In Sec.\ \ref{sec:feedback-dynamics} we investigate the phases of fluctuations around the critical temperature. 

\section{\label{sec:model}Model description}
We describe the decision of a single agent to buy or to sell using the GRVM as a typical opinion formation model \cite{krause12}. $N=L^2$ agents on a two dimensional lattice with periodic boundary conditions adapt their opinions $s_i=\pm 1$ (buy or sell) in random sequential update to their four nearest neighbors (denoted by nn$(i)$). Using the number of agreeing neighbors 
\begin{align}
    u_i=\sum_{j \in {\rm nn}(i)}\delta_{s_i,s_j},
\end{align}
the system's dynamics is described with the flip-probabilities $p_{u\to 4-u}$ (we set $p_{u\to 4-u}+p_{4-u\to u}=1$). With suppressed voluntary isolation we get $p_{0 \to 4}=1$, $p_{2 \to 2}=1/2$ and $p_{4 \to 0}=0$. The remaining probability of joining local minorities $p_{3\to 1}$ is set with the inverse temperature $\beta=1/T$ to
\begin{align}
    p_{3 \to 1} &= \frac{1}{1+\exp(4 \beta)} & p_{1 \to 3}  &=  1-p_{3 \to 1},\label{eq:switch}  
\end{align}
as in heat bath Monte Carlo simulations of the Ising model. 

\begin{figure}[htb]
\begin{center}
	\includegraphics[width=0.7\columnwidth]{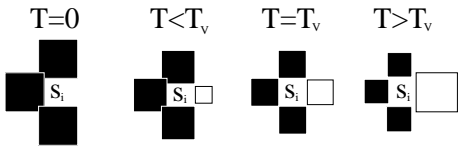}
	\caption{Illustration of the persuasiveness of the neighbors of an agent $s_i$ for different temperatures $T$. For low temperatures, majorities convince more strongly. For high temperatures, the agent follows any present opinion in a panic-like mood.}
		\label{fig:persuasive}
\end{center}
\end{figure}
For low temperatures we have an increased persuasiveness of local groups of agents ($p_{1 \rightarrow 3} > 3 p_{3 \rightarrow 1}$), as illustrated in Fig.\ \ref{fig:persuasive}. For $T = T_{\rm V} = 4/\ln(3) \approx 3.641$ we get the voter model, because for this temperature every neighbor convinces with the same probability ($p_{3 \to 1}=1/4$ and $p_{1 \to 3}=3/4$). For higher temperatures $T>T_{\rm V}$, local majorities have a suppressed persuasive power. Agents adopt any opinion in their neighborhood without trusting local majorities which can be seen as a panic-like behavior. The social temperature $T$ can be seen as the market temperature because it influences the uncertainty of investment strategies of all agents. In the next section we will see that the market temperature is connected to the market's behavior and volatility in an intuitive way. 

Agents buy or sell one unit every time step (sweep). Therefore the opinion polarization 
\begin{align}
m &= \frac{1}{N}\sum_j s_j
\end{align}
translates into the market imbalance. Market imbalances have two consequences: They are at the basis of the pricing process for our artificial market, and they are the central ingredient for the introduction of the feedback.

Assuming a constant fundamental value of the asset, price changes are related to the market imbalance using logarithmic returns as ${\rm ret}(t)=\ln(p(t)/p(t-1))\propto\Delta m=m(t)-m(t-1)$. This can be motivated using additional fundamentalists taking advantage of arbitrage possibilities in balance with their risk \cite{kaizoji2002}. As a volatility measure, $\sigma=(\left<\Delta m^2\right>-\left<\Delta m\right>^2)^{1/2}$ with averages over short sequences or whole time series can be used. 

Let us now introduce a feedback relating increasing market imbalances to an increase in market temperature in the form 
\begin{align}
T &= T(|m|) \quad \textrm{monotone}\label{eq:feedback}\\
T(0) &< T_{\rm V} < T(1).
\end{align}
This feedback is consecutively updated with the systems dynamics. For a balanced market, agents trust in local majorities (as is the case for $T(0)<T_{\rm V}$), while for high market imbalances any present opinion is adopted in a panic like fashion ($T(1)>T_{\rm V}$). The idea to introduce a feedback of the described form is taken from \cite{bornholdt2001}, where $m$ controls the switching probabilities through an additional term in the Ising model. In Appendix~\ref{app:spinmarket} the market temperature for this model is estimated.

\section{\label{sec:constant}Constant market temperature}
It is instructive to analyze the dynamics for constant temperature. With this we analyze different market modes contributing to the dynamics of the model with feedback. By doing so we connect to the opinion formation background. This will help us to understand the feedback mechanism in Sec.~\ref{sec:feedback}~and~\ref{sec:feedback-dynamics}. 

\begin{figure}[htb]
\begin{center}
	\includegraphics[width=1.0\columnwidth]{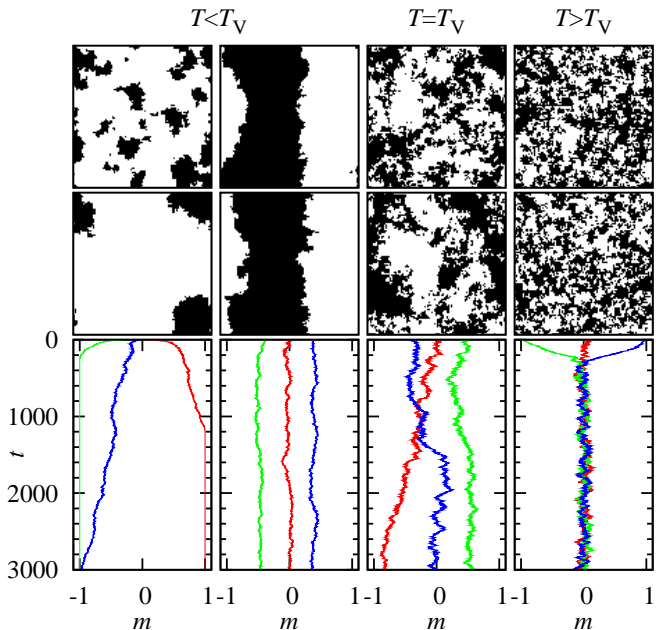}
	\caption{Typical system states (top) and typical trajectories of $m$ (bottom, second column from left with striped initial states, all others with biased random initial conditions, different initial imbalances $m(t=0)$) for $T=0.6\,T_{\rm V}$ (first and second column from left), $T=T_{\rm V}$  and $T=1.2\,T_{\rm V}$. Subcritical temperatures lead to polarization or metastable striped states, the critical lattice configurations show clusters of diverse sizes and the supercritical market is balanced but volatile.}
    \label{fig:grvm}
\end{center}
\end{figure}
Fig.~\ref{fig:grvm} shows some system states (top) and opinion polarization time series (bottom) for $T=0.6\,T_{\rm V}$ (first and second column from left), $T=T_{\rm V}$ (third column) and $T=1.2\,T_{\rm V}$ (fourth column). In the first column with a subcritical temperature, where local majorities are strongly preferred and thus single agents follow local group opinions, we see ordering dynamics leading to absorbing consensus states. The three shown time series with biased random initial conditions and magnetizations about $m=-0.2$, $m=-0.04$ and $m=0.1$ experience a strong drift away from the balanced market which grows with the initial market imbalance. This scenario leads to a totally dysfunctional market. 

In the second column a subcritical system with metastable striped states can be seen. As discussed in Appendix \ref{app:coarsening}, such states can emerge for random initial conditions with opinion polarization around $m=0$, as known for similar opinion formation models \cite{chen2005}, but also from strongly polarized states with global structures. The system reaching metastable states is not forced to market imbalances and thus can preserve low market imbalances while performing a random walk with low volatility ($\sigma=0.003$ 
for the shown time series). As typical for random walks, bull and bear market durations are power law distributed (here with cut-off) and thus market imbalances survive for long times. Metastable states finally end up in an absorbing state and have a typical life-time dependent on temperature and system size (about $40\,000$ sweeps for the shown case). 

In the third column, the critical temperature of the model leads to a drift-free behavior identical to the VM. The volatility depends on the opinion polarization with maximum values about $\sigma=0.008$ around $m=0$. This higher volatility leads to shorter times before absorbing states are reached. The system states (for infinite systems and infinite coarsening times) show clusters of all length scales and thus in the finite system global structures emerge. Finally the fourth column shows a disordered system with strong drift forces leading to a balanced market situation. In contrast to the subcritical metastable states, volatility is high ($\sigma=0.010$).

In conclusion, the effects of the market temperature are intuitive in several respects: Single agents notice a faster changing environment, change their own opinion more frequently and exhibit only limited trust in local majorities for higher temperatures. The returns as a global property of the system increase with temperature rendering the system more volatile.

\section{\label{sec:feedback}Market with feedback}
We analyze an artificial market with $N=200^2$ agents and a feedback
\begin{align}
T(|m|)/T_{\rm V} &= 0.2+\alpha\cdot |m|\label{eq:feedback1}
\end{align}
with feedback strength $\alpha=10$. This kind of linear feedback is a particularly simple variant of Eq.~(\ref{eq:feedback}) with a temperature above zero at $m=0$. We will see, that imbalances with the critical temperature $T_{\rm V}$ are of special interest in describing the dynamics, therefore we define $m_{\rm F}$ due to $T(\pm m_{\rm F})=T_{\rm V}$ with the specific value $m_{\rm F} = 0.08$ for this feedback. 

\begin{figure}[htb]
\begin{center}
	\includegraphics[width=1\columnwidth]{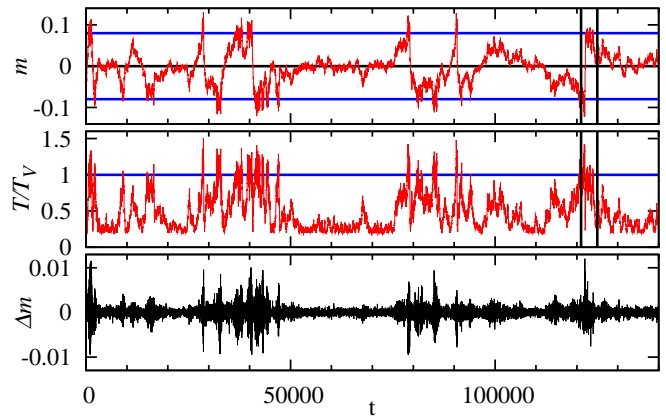}
	\caption{Time series of the market imbalance $m$ with values $\pm m_{\rm F}$ (top), market temperature $T$ (middle) and returns $\Delta m$ (bottom) for a system with $N=200^2$ and the feedback of Eq.\ (\ref{eq:feedback1}). 
	The returns span a wide range and appear clustered, as typical for price time series.}
    \label{fig:meta_excursions}
\end{center}
\end{figure}
In Fig.\ \ref{fig:meta_excursions} time series of the market imbalance, market temperature and returns are shown. All simulations are started from random initial conditions. 
We see long periods of low market imbalance $m$, low market temperature $T$ and low volatility (small returns $\Delta m$, note especially the period between $t\approx 50000$ and $t\approx 70000$) alternating with short periods of fluctuations around the critical temperature $T_{\rm V}$ connected to high imbalances around $m_{\rm F}$ and high volatility (as for example the time window indicated with vertical lines in the upper two rows). Considering the whole time series, returns are broadly distributed (see Fig.\ \ref{fig:rho_feedbacks}) and clearly occur clustered, as known from real markets. Changing temperatures are an intuitive property of real markets. In this artificial market, market imbalances (representing speculative risk) lead to nervous reactions of agents incorporated with higher market temperature, culminating in panic like behavior above $T=T_{\rm V}$. 

During long periods of low market imbalance, the system shows striped lattice states as in Fig.~\ref{fig:grvm}, second column from left. In these states, market imbalance performs a drift-free diffusion due to diffusing border lines between the two stripes. As in such states the system stays for long times near to the fundamental value, the market is closest to the case of rational agents (independent agents using adequate but noisy information about the fundamental value to maximize their outcome), but with the tendency to large imbalances due to undirected diffusion. Bull and bear market durations are power law distributed (with a cutoff) as they are for the market with constant temperature showing striped lattice states. 
If the market being in striped states reaches large imbalances leading to panic-like reactions of agents, fluctuations around the critical temperature help to find a balanced market with striped states within short times (compared to typical time scales of the market with striped states). This is surprising, since single agents in our model do not act in a purposeful way. 

Both, periods with striped states and periods of fluctuations around the critical temperature, represent metastable phases, as their typical time scales diverge with system size. We performed simulations (not shown) with $N=100^2$ ($5\cdot 10^5$ sweeps) up to $N=800^2$ ($32\cdot10^6$ sweeps) using a scaling for the feedback strength $\alpha\propto L^{1/2}$ (reflecting the adjustment of typical coarsening times on the way to striped states $\propto L^2$ and typical times in striped states $\propto L^3/\alpha^2$, see \cite{krause2011}). We found a similar behavior of the system for different system sizes taking into account that typical time scales grow with system size $N=L^2$. For the infinite system thus feedback strength as well as time scales diverge. This emphasizes the finite-size character of the system in accordance with the finite-size-effect of striped states.

\section{\label{sec:feedback-dynamics}Dynamic feedback mechanism}
\subsection{\label{sec:fluctuations}Fluctuations around fixed points}
In the last section we have seen fluctuations of the market imbalance $m$ around values $m_{\rm F}$ connected to the critical temperature, $T(m_{\rm F})=T_{\rm V}$. To get a better understanding of this behavior, let us use a Langevin equation for the mean-field behavior. We transfer a Fokker-Planck description of the GRVM \cite{krause12} including drift and diffusion effects to the description of trajectories ($\Delta t=1$, $\Delta m(t)=m(t+1)-m(t)$, see also \cite{cast09,ham05,vazquez2008} for the discussion of the Langevin approach for nonlinear voter models),
\begin{align}
\Delta m(t) &= a_1(m(t),T)+\sqrt{a_2(m(t))}\cdot \xi_t \label{eq:langevin},\\
a_1(m,T) &= -\phi_1\cdot 2\left(p_{3 \to 1}(T)-\frac{1}{4}\right)m(1-m^2)\label{eq:drift},\\
a_2(m) &= \phi_2\frac{2}{N}(1-m^2)\label{eq:diffusion},
\end{align}
with an iid.\ random variable $\xi_t$ being normally distributed and $p_{3\to 1}$ of Eq.~(\ref{eq:switch}). The parameters take the values $\phi_1 = \phi_2 = 1$ for the mean-field case. The 2D lattice behaves mean-field--like also with values $\phi_1, \phi_2 < 1$, as has been found with numerical measurements of drift and diffusion \cite{krause12}. 

\begin{figure}[htb]
\begin{center}
	\includegraphics[width=1.0\columnwidth]{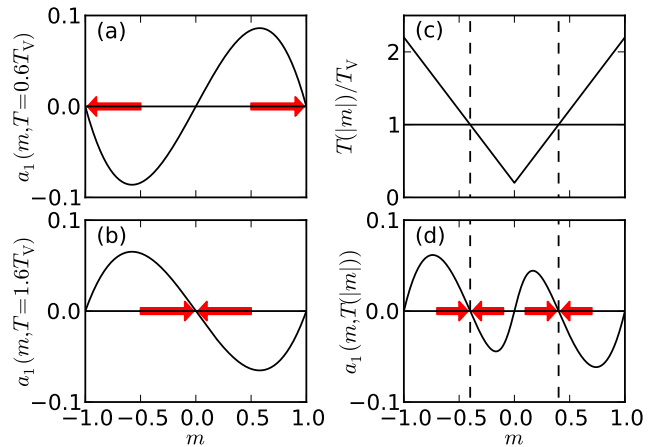}
	\caption{Mean-field drift term Eq.~(\ref{eq:drift}) for a constant subcritical (a) and supercritical temperature (b) driving the system to fully polarized absorbing states(a), respectively to a non-polarized stable fixed point (b), as indicated by arrows. Inbetween at $T=T_{\rm V}$ the system is drift-free.  With the feedback $T(|m|)$ (c), the additional implicit $m$-dependence of the drift leads to two stable fixed points at intermediate polarization $\pm m_{\rm F}$ (d) connected to the critical temperature $T_{\rm V}$. With non-zero diffusion Eq.~(\ref{eq:diffusion}), finite size systems perform fluctuations around the fixed points.}
    \label{fig:drift}
\end{center}
\end{figure}
First of all we ignore fluctuations due to finite diffusion (Eq.~(\ref{eq:diffusion})) and have a look at the fixed points connected to drift forces (Eq.~(\ref{eq:drift})) in the deterministic limit $a_2(m)\to 0$. In Fig.~\ref{fig:drift} on the left, the drift term $a_1(m,T)$ is shown for two different constant temperatures. We see that a subcritical temperature leads to a drift in direction of total polarization (a), indicated with arrows (compare Fig.~\ref{fig:grvm} outer left). In the supercritical case (b), drift tends to reduce polarization (compare Fig.~\ref{fig:grvm} outer right). With the feedback (c) (here Eq.~(\ref{eq:feedback1}) with $\alpha=2$), the drift term $a_1(m,T(|m|))$ gets an additional implicit dependency on $m$ via $T$ (d). The drift forces lead to two fixed points (provided the feedback is strictly monotone at these points): 
\begin{align}
m &= \pm m_{\rm F} \quad \textrm{fixed points}\\
T(m_{\rm F}) &= T_{\rm V}.\label{eq:critical_fp}
\end{align}
The value $m_{\rm F}$ is defined according to $T(|m|<m_{\rm F})<T_{\rm V}$ and $T(|m|>m_{\rm F})>T_{\rm V}$. Eq.\ (\ref{eq:critical_fp}) is only valid, if $T(|m|)$ is continuous around the fixed points. A counter example is discussed below. With the balancing drift forces for high imbalances $m$, absorbing states with total consensus are strongly suppressed. 

\begin{figure}[htb]
\begin{center}
	\includegraphics[width=1\columnwidth]{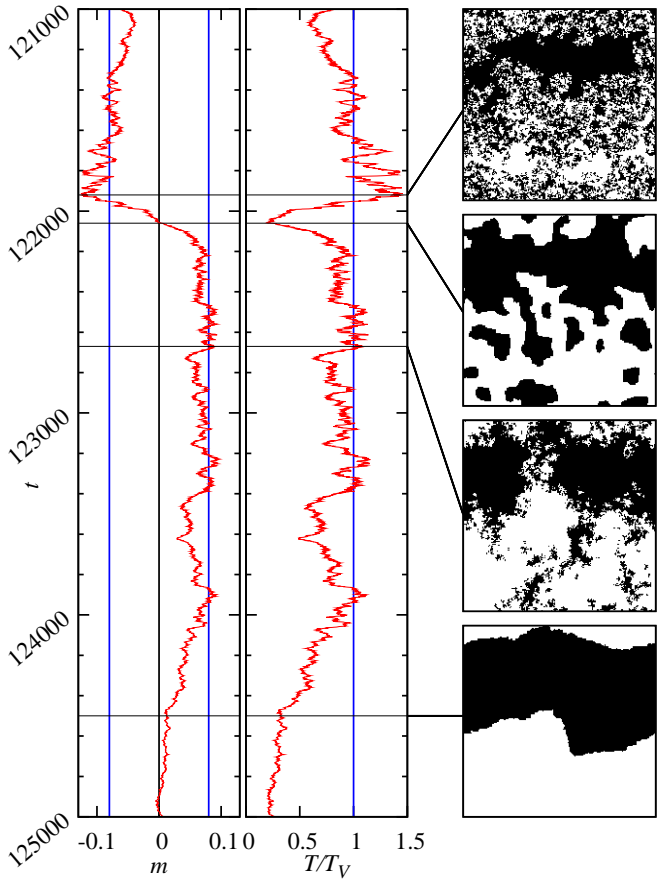}
	\caption{Time series of Fig.\ \ref{fig:meta_excursions} (for the time window indicated there) shown in rotated orientation (left and center). For the times indicated by horizontal lines, the according system states are shown (right). Large fluctuations around the fixed points $m_{\rm F}$ are in accordance with Eq.~(\ref{eq:langevin}) and lead to successive melting and coarsening processes. The striped state at the end without drift forces allows for a long excursion away from the fixed point.}
    \label{fig:dynamics_2}
\end{center}
\end{figure}
For a system of finite size, $m$ fluctuates around the fixed points according to the interplay of $a_1$ and $a_2$. With Fig.\ \ref{fig:dynamics_2} we see such fixed point fluctuations for the system discussed in the last section and the short time window as indicated in Fig.\ \ref{fig:meta_excursions}. The time series (shown with rotated orientation compared to Fig.~\ref{fig:meta_excursions}) show fixed point fluctuations between $t=121300$ and $t=123400$. For this time interval, the time series of $m$ and $T$ are compatible with Eq.~(\ref{eq:langevin}) concerning time scales and amplitudes of fluctuations as well as occasional switches between the two fixed points (we used the estimation $\phi_1=0.1$ and $\phi_2=0.3$ taken from numeric measurements in \cite{krause12} and generated a mean-field time series, not shown). Therefore, fluctuations around the fixed points for the 2D system can partly be understood to be mean-field like. This is remarkable in respect to strong correlations of the lattice states, as shown with the upper three lattice states on the right of the figure for times indicated with horizontal lines around the time series. As the fluctuations span a wide range of temperature $T$, consecutive melting and coarsening processes can be seen. In general, correlations limit quantitative predictions of Eq.~(\ref{eq:langevin}). Accordingly fluctuations with long times until return to $T_{\rm V}$, as for example between $t=123400$ and $t=124000$, are over-represented compared to the mean-field time series, where fluctuations with more than 200 time steps only account for two percent of the total time.

At the end of the shown time window, the system is in striped lattice states again (see the lattice state at the bottom of Fig.\ \ref{fig:dynamics_2}). This leads to another long period of small $m$ far away from the fixed points (compare Fig.~\ref{fig:meta_excursions}). By generating long range correlations during the coarsening periods which are changed fast during melting periods, the system self-organizes to striped states during fixed point fluctuations. This process will be analyzed in the next subsection for different feedbacks confirming its robustness.

\subsection{Robustness of the feedback process}

\begin{figure}[htb]
\begin{center}
	\includegraphics[width=1\columnwidth]{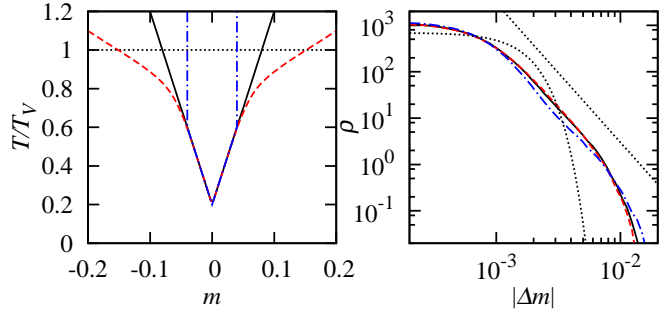}
	\caption{Return densities (right) for the different feedback variants (shown left) of Eq.\ (\ref{eq:feedback1}) (black line), Eq.\ (\ref{eq:feedback2}) (blue dash-dotted line) and Eq.\ (\ref{eq:feedback3}) (red dashed line). Although the lattice states differ largely, the outcome (as well as the time series) is quite similar indicating a sign of robustness. On the right, a Gaussian distribution with the same variance and a power law with exponent $-3$ are shown for comparison.}
    \label{fig:rho_feedbacks}
\end{center}
\end{figure}
In this subsection we vary the feedback as shown in Fig.~\ref{fig:rho_feedbacks} on the left and find differing behavior concerning the system states, nonetheless with weak effects on the global behavior. To demonstrate the effect of consecutive coarsening and melting periods, we use an extreme feedback 
\begin{align}
T(|m|)/T_{\rm V} &=\begin{cases}
         0.2+\alpha\cdot |m|, & |m|\leq m_{\rm F}\\
         1000, & |m|>m_{\rm F}
       \end{cases}.\label{eq:feedback2}
\end{align}
with $\alpha=10$ and $m_{\rm F}=0.04$ leading to a jump at $T=0.6\,T_{\rm V}$ (system size $N=200^2$). The temporal behavior concerning the time series as well as the lattice states does not change considerably compared to the case described before. Therefore the process of consecutive melting and coarsening processes is clearly confirmed. The broad distribution of returns is hardly changed, as can be seen in Fig.\ \ref{fig:rho_feedbacks} on the right. 

For comparison, let us choose a feedback 
\begin{align}
T(|m|)/T_{\rm V} &=0.2+\left[ \frac{1}{(\alpha |m|)^{6}}+\frac{1}{\left(0.5+2|m|\right)^{6}} \right]^{-\frac{1}{6}}\label{eq:feedback3}
\end{align}
with $\alpha=10$ which is smoother near the fixed points. In combination with higher drift forces due to fixed points at about $m_{\rm F} = \pm 0.1515$ (the maximum of the drift forces lies at $m=0.5$), this leads to weak fluctuations of the temperature around the critical value $T_{\rm V}$. 

\begin{figure}[htb]
\begin{center}
	\includegraphics[width=1\columnwidth]{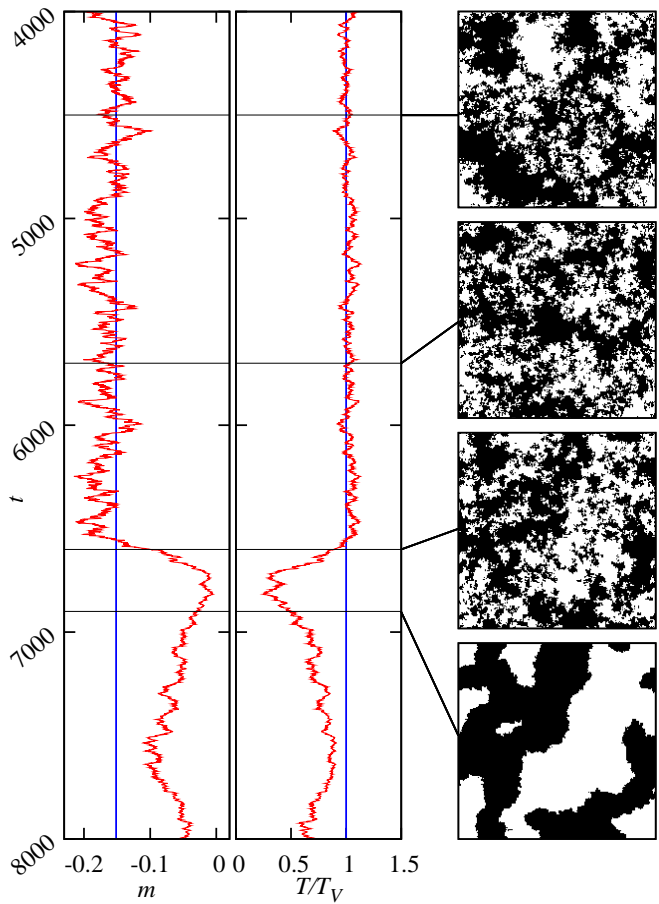}
	\caption{The same as in Fig.\ \ref{fig:dynamics_2} for the smooth feedback of Eq.\ (\ref{eq:feedback3}). Little fluctuations around the fixed point lead to similar states as for the voter model. Finally a coarsening process leads to striped states.}
    \label{fig:dynamics_smooth}
\end{center}
\end{figure}
In Fig.\ \ref{fig:dynamics_smooth} a long period of fixed point fluctuations is followed by a system with striped states. The corresponding lattice states during the fixed point fluctuations are comparable to critical states (compare Fig.\ \ref{fig:grvm} third column from left), the last snapshot shows the coarsening process on the way to striped states. The lattice states near the fixed points are quite interesting, since they show similarity to the states of the voter model, but with a pronounced opinion polarization. Fluctuations in direction of lower temperatures sometimes lead to a balancing coarsening process. The coarsening system starting from a state with large clusters can lead to striped states even for large initial polarization (see Appendix \ref{app:coarsening}). Again, the return density is hardly changed compared to the feedback of Eq.\ (\ref{eq:feedback1}) (see Fig.\ \ref{fig:rho_feedbacks}), although the lattice states during the self-organization to striped 
 states look quite different. 

As we have seen, the self-organization to striped states is a robust process, since it works for large ranges of parameters. Nonetheless, feedbacks of more seriously differing shapes lead to a variety of effects. As an example, a system with a steep feedback at large $m_{\rm F}$ can be trapped in a structured state as, for example, a circle which can hardly be escaped during repeated fixed point fluctuations.

\begin{figure}[htb]
\begin{center}
	\includegraphics[width=0.6\columnwidth]{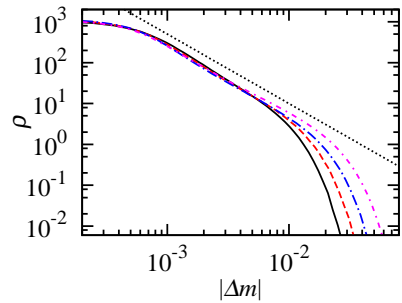}
	\caption{Return densities for the Ising model and linear feedbacks for $N=L^2=128^2$, $N=256^2$, $N=512^2$ and $N=1024^2$ (distributions of increasing width, rescaled by a factor $(L/128)^{3/2}$). For comparison the dotted line shows a power law with exponent $-1.7$.}
    \label{fig:rho_ising}
\end{center}
\end{figure}
The feedback process shows another sign of robustness, since it works in other systems as well. We here exemplify this using the pure Ising model with heat bath dynamics and a feedback  
\begin{align}
T(|m|)/T_{\rm C}^{\rm Ising} &= 0.2+\alpha_{\rm Ising}\cdot |m|\label{eq:feedback4}
\end{align}
for $L=128$ with $\alpha_{\rm Ising}=10$ and $10^6$ sweeps up to $L=1024$ with $\alpha_{\rm Ising}=28$ and $64\cdot10^6$ sweeps using the scaling $\alpha\propto L^{1/2}$ and ${\rm time}\propto L^2$. We find time series close to Fig.\ \ref{fig:meta_excursions}. As can be seen in Fig.\ \ref{fig:rho_ising}, the returns are broadly distributed with a flatter decline compared to the opinion formation system. 

\section{Summary}

We translate a herding market model on a 2D lattice with feedback \cite{bornholdt2001} to an opinion formation model for two reasons. First this helps us to find an intuitive insight with a market temperature controlled by market imbalance. Second we can use the framework of opinion formation to get a better understanding of the dynamic mechanisms ruling this model class. 

We introduce the GRVM as an implementation for the herding process with constant temperature. As a manifestation of fear, each agent is coupled to an increasing market temperature for increasing market imbalances. In this way a feedback is introduced. 
For constant market temperature, we describe different market modes and confirm the intuitive effects of the market temperature. 

The market with feedback exhibits two kinds of metastable phases (as their typical times increase with system size). Short phases with mean-field-like fluctuations around critical fixed points of large market imbalance appear as a robust self-organization to striped lattice states. This means that single agents with their panic-like behavior enable the system to find ordered states. Long phases with striped states lead to undirected diffusive movements of the market imbalance (as described in detail in \cite{krause2011}) with low market imbalance and low volatility and are close to the efficient market. 

Future work could quantify the behavior near the fixed points by 
performing Langevin simulations and using probabilities for striped states in a coarsening process started with correlated lattice states. 

While the focus of this article is the limit of a regular 2D lattice, it would be interesting to introduce a number of shortcut links between pairs of randomly chosen agents. As fluctuations around the critical fixed points are mean-field-like, they are likely similar for a higher effective dimension of the system. On the other hand, correlated states in a small world network should be more diverse. Besides a communication structure of agents being closer to the case of real markets, this could cause a more diverse memory of the system with an auto correlation of absolute returns closer to real markets.

Finally, it is worth comparing the model to advanced opinion formation models beyond simple consensus formation, including further effects of real social systems. In \cite{nyczka2012}, different kinds of nonconformity as the existence of contrarians \cite{galam2004} are investigated as mechanisms preventing consensus. The degree of noncornformity there can be connected to a constant social temperature leading to different kinds of phase transitions. Systems in the disordered phase including a noisy global feedback are used to describe financial markets \cite{krawiecki2002,bartolozzi2005}. The obvious lack of consensus in elections with typical small differences between winner and second place can be described with a potts model variant in the ordered phase including a global feedback weakening the leading candidate in polls \cite{araujo2010}. This feedback selectively prefers single opinions and therefore does not act as an effective increase of temperature. In the present article another mechanism preventing total agreement is presented for a system being in the ordered phase most of the time. With the global feedback on temperature, crises (with panic-like behavior as a variety of nonconformity) help to restore the balance of the system.

\appendix

\section{}\label{app:spinmarket}
At this place we want to estimate the effective temperature caused by the feedback in the non-equilibrium model of \cite{bornholdt2001}. 
In our notation, with the number of agreeing neighbors $u$, the switching probabilities of this model read 
\begin{align}
{\tilde p}_{u\to (4-u)}(|m|) &= \frac{1}{1+\exp[\tilde{\beta}_0 (4u-8-2\tilde{\alpha}|m|)]}
\end{align}
with the parameters $\tilde{\beta}_0$ (inverse temperature of the Ising model for $m=0$) and $\tilde{\alpha}$ (coupling parameter). For a low temperature around $\tilde{T}_0=0.5$ and accordingly $\tilde{\beta}_0=1/\tilde{T}_0=2$, the switching probabilities realized by the system lead to a behavior close to the case of the GRVM 
with switching probabilities $\tilde p_{0\to 4}\approx 1$ and $\tilde p_{4\to 0}\approx 0$. 
Consequently the ratio of the switching probabilities 
${\tilde p}_{3\to 1}/{\tilde p}_{1\to 3}$ determines the effective temperature of the system 
(for fundamental considerations on the concept of temperature in non-equilibrium systems, exemplified using kinetic Ising models, see \cite{sastre2003}). Starting from $p_{3\to 1}(T)/p_{1\to 3}(T)={\tilde p}_{3\to 1}(|m|)/{\tilde p}_{1\to 3}(|m|)$, with an additional approximation we get 
\begin{align}
T(|m|) &\approx \frac{4}{\ln\{1+\exp[\tilde{\beta}_0(4-2\tilde{\alpha}|m|)]\}}.
\end{align}
This is a monotone function starting at $T(0)\approx \tilde T_0$ and reaching the critical value $T_{\rm V}$ at about $|m|=1.84/\tilde{\alpha}$ ($\tilde\alpha$ is typically chosen $\tilde{\alpha}\geq 10$). As the critical value is not exceeded very much by $m$ \cite{krause2011}, $\tilde p_{4\to 0}\approx 0$ is justified ($\tilde p_{4\to 0}\approx 0.00007$ at the critical $|m|$ \cite{krause2011} which is small compared for example to the critical Ising model with $p_{4\to 0}\approx 0.0286$).

\section{}\label{app:coarsening}
We here are interested in striped states which are essential for the system with feedback. Such states are known to occur during the coarsening process as reported for similar opinion formation models of finite size \cite{chen2005}. The coarsening dynamics of the GRVM is well known in terms of the temporal behavior of cluster sizes and the interfacial density \cite{dorn01}.

\begin{figure}[htb]
\begin{center}
	\includegraphics[width=1.0\columnwidth]{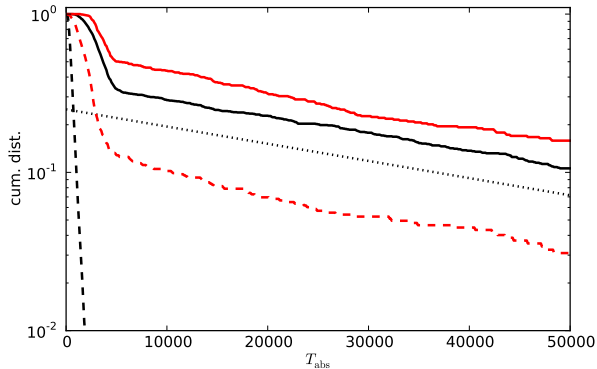}
	\caption{Cumulative distribution of the consensus times $T_{\rm abs}$ with $N=100^2$ and $T=0.6\,T_{\rm V}$ for random initial conditions  (black solid line), biased random initial conditions with $m\approx0.14$ (black dashed line) and starting from critical lattice states with $m<0.14$ (red solid line) and $m>0.14$ (red dashed line). For comparison, $0.25\exp(-T_{\rm abs}/\tau)$ with $\tau=40000$ is plotted (dotted line). The fast coarsening process for short times leads to metastable striped states with different branching ratios.}
    \label{fig:cum_t_abs}
\end{center}
\end{figure}
In Fig.\ \ref{fig:cum_t_abs} cumulative distributions of consensus times $T_{\rm abs}$ are shown for different classes of initial conditions and $N=100^2$, $T=0.6\,T_{\rm V}$. Biased random initial conditions with high opinion polarization around $m=0.14$ lead to a fast consensus with probabilities of reaching metastable striped states in the sub-percent regime (dashed black line). In contrast random initial conditions around $m=0$ lead to a metastable striped state in about every third case (black solid line). The metastable striped states show typical lifetimes of about $\tau=40000$, which grows with system size and shrinks with temperature. 

The red lines show distributions for critical states (showing more structure than random initial conditions) as initial states. Random initial conditions were propagated $1000$ sweeps at $T=T_{\rm V}$ and at $t=0$ the temperature was quenched to $T=0.6\,T_{\rm V}$. For little opinion polarization at $t=0$, $m<0.14$, this leads to metastable striped states in about every second case (red solid line). Highly polarized states with $m>0.14$ show a smaller but nevertheless significant branching ratio (red dashed line).

\begin{acknowledgments}
We acknowledge the support of the DFG under contract No. INST 144/242-1 FUGG. 
\end{acknowledgments}


\begin{thebibliography}{00}

\bibitem{sornette2004}
D.\ Sornette, {\sl Why Stock Markets Crash: Critical Events in Complex Financial Systems}, (Princeton University Press, Princeton, 2004).

\bibitem{castellano2009}
C.\ Castellano, S.\ Fortunato, and V.\ Loreto, 
Rev. Mod. Phys.\ 81, 591-646 (2009).

\bibitem{krause2011}
S.M.\ Krause and S.\ Bornholdt, 
Spin models as microfoundation of macroscopic financial market models, 
arXiv:1103.5345v1; submitted. 

\bibitem{bornholdt2001}
S.\ Bornholdt, 
Int. J. Mod. Phys. C 12, 667-674 (2001).

\bibitem{kaizoji2002}
T.\ Kaizoji, S.\ Bornholdt, and Y.\ Fujiwara, 
Physica A 316, 441-452 (2002).

\bibitem{yamamoto2010}
R.\ Yamamoto, 
Physica A 389, 1208-1214 (2010).

\bibitem{mandelbrot1963}
B.\ Mandelbrot, 
J.\ Business 36, 394-419 (1963).

\bibitem{gopikrishnan1999}
P.\ Gopikrishnan, V.\ Plerou, L.A.\ Nunes Amaral, M.\ Meyer, and H.E.\ Stanley, 
Phys.\ Rev.\ E 60, 5305-5316 (1999).

\bibitem{clifford1973}
P.\ Clifford and A.\ Sudbury, 
Biometrika  60, 581-588 (1973).

\bibitem{dorn01}
I.\ Dornic, H.\ Chat\'e, J.\ Chave, and H.\ Hinrichsen, 
Phys. Rev. Let.\ 87, 045701(2001).

\bibitem{hin10}
H.\ Hinrichsen, 
Advances in Physics 49, 815-958 (2000).

\bibitem{odor04}
G.\ \'Odor, 
Rev. Mod. Phys. 76, 663 (2004).

\bibitem{ham05}
O.\ Al Hammal, H.\ Chat\'e, I.\ Dornic, and M.A.\ Mun\~oz, 
Phys. Rev. Let. 94, 230601 (2005).

\bibitem{cast09}
C.\ Castellano, M.A.\ Mu\~noz, and R.\ Pastor-Satorras, 
Phys. Rev. E 80, 041129 (2009).

\bibitem{vazquez2008}
F.\ Vazquez and C.\ L\'opez, 
Phys. Rev. E 78, 061127 (2008).

\bibitem{krause12}
S.M.\ Krause, F.\ B\"ottcher, and S.\ Bornholdt, 
Phys. Rev. E 85, 031126 (2012).

\bibitem{chen2005}
P.\ Chen and S.\ Redner, 
Phys.\ Rev.\ E 71, 036101 (2005).

\bibitem{bordogna2011}
C.M.\ Bordogna and E.V.\ Albano,
Phys. Rev. E 83, 046111 (2011).

\bibitem{stark2008}
H.-U.\ Stark, C.J.\ Tessone, and F.\ Schweitzer, 
Phys. Rev. Lett 101, 018701 (2008).

\bibitem{oliv93}
M.J.\ de Oliveria, J.F.F.\ Mendes, and M.A.\ Santos, 
J. Phys. A 26, 2317 (1993). 

\bibitem{nyczka2012}
P.\ Nyczka, K.\ Sznajd-Weron, and J.\ Cis\l{}o
Phys. Rev. E 86, 011105 (2012).

\bibitem{galam2004}
S.\ Galam, 
Physica A 333, 453–460 (2004).

\bibitem{krawiecki2002}
A.\ Krawiecki, J.A.\ Ho\l{}yst,  and D.\ Helbing, 
Phys. Rev. Lett. 89, 158701 (2002).

\bibitem{bartolozzi2005}
M.\ Bartolozzi, D.B.\ Leinweber, and A.W.\ Thomas, 
Phys. Rev. E 72, 046113 (2005).

\bibitem{araujo2010}
N.A.M.\ Ara\'ujo, J.S.\ Andrade Jr, and H.J.\ Herrmann,
PLoS ONE 5, e12446 (2010).

\bibitem{sastre2003}
F.\ Sastre, I. Dornic, and Hugues Chat\'e, 
Phys. Rev. Lett. 91, 267205 (2003).

\end{thebibliography}
\end{document}